\documentclass{article}
\usepackage{spconf,amsmath,graphicx}
\usepackage{booktabs}
\usepackage{multirow}
\usepackage{array}
\usepackage{url}
\usepackage{cite}
\usepackage{color, colortbl}
\usepackage{xcolor}
\colorlet{top5}{blue!10}
\colorlet{sig}{orange!10}
\colorlet{high_cor}{red!10}

\usepackage{makecell}

\newcommand*\samethanks[1][\value{footnote}]{\footnotemark[#1]}

\title{The VoiceMOS Challenge 2023:\\Zero-shot Subjective Speech Quality Prediction for Multiple Domains}
\name{
Erica Cooper$^{1}$\sthanks{Equal contribution.}, Wen-Chin Huang$^{2}$\samethanks[1], Yu Tsao$^3$, Hsin-Min Wang$^3$, Tomoki Toda$^2$, Junichi Yamagishi$^1$
}
\address{
  $^1$National Institute of Informatics, Japan\\
  $^2$Nagoya University, Japan\\
  $^3$Academia Sinica, Taiwan
}
\copyrightnotice{979-8-3503-0689-7/23/\$31.00~\copyright2023 IEEE}
\begin{document}
\ninept
\maketitle
\begin{abstract}
We present the second edition of the VoiceMOS Challenge, a scientific event that aims to promote the study of automatic prediction of the mean opinion score (MOS) of synthesized and processed speech. This year, we emphasize real-world and challenging zero-shot out-of-domain MOS prediction with three tracks for three different voice evaluation scenarios.  Ten teams from industry and academia in seven different countries participated.  Surprisingly, we found that the two sub-tracks of French text-to-speech synthesis had large differences in their predictability, and that singing voice-converted samples were not as difficult to predict as we had expected.  Use of diverse datasets and listener information during training appeared to be successful approaches.  

\end{abstract}

\section{Introduction}
\label{sec:intro}

In 2022, the first edition of the  VoiceMOS challenge \cite{voicemos2022} was organized to encourage research in the area of automatic quality assessment for synthesized speech.  The task of the challenge was to predict Mean Opinion Score (MOS) ratings given audio samples.  Participants were provided with a large-scale MOS dataset \cite{bvcc} containing text-to-speech and voice conversion samples spanning research and development efforts from 2008-2020 with some MOS labels withheld for evaluation.  Participants could also participate in a second, optional out-of-domain (OOD) track where a much smaller amount of labeled data was provided to participants from a separate listening test.  In particular, with only 136 labeled samples, a system-level Spearman rank correlation coefficient (SRCC) of 0.979 was obtained by the top system.  The challenge succeeded in drawing attention and interest towards this growing research area, and the results demonstrated that MOS can be well-predicted when some in-domain listening test data is already available.

However, the results of last year's challenge revealed that MOS ratings for samples from previously-unseen speakers and synthesis systems tend to be more difficult to predict.  Indeed, these are important scenarios because they reflect real-life MOS prediction -- researchers developing new synthesis systems will by definition be creating unseen systems, and they may be working with data from different speakers or even languages from what was seen in the training data of a MOS predictor.  Many participants also showed interest in subjective quality prediction of other speech types such as noisy and enhanced speech.  While the speech enhancement community has long relied on the full-reference objective metrics like PESQ (perceptual evaluation of speech quality), there has been a recent trend on non-intrusive speech quality assessment (NISQA) \cite{nisqa}, and a challenge similar to VoiceMOS also took place in parallel in 2022 \cite{conferencingspeech2022}.  This raises the question of whether \textit{a single model can predict the perceptual quality scores of speech samples from multiple domains}.  

With these motivations in mind, the 2023 edition of the VoiceMOS Challenge focused on {\em real-world and challenging zero-shot out-of-domain MOS prediction.}  Participants were presented with three very different tracks in which to participate, and they were not provided with any MOS-labeled data in two of those domains, as MOS ratings had not in fact yet been collected at the time that the audio samples were distributed.  The challenge therefore reflects a real-life MOS prediction scenario.

For this year's challenge, in collaboration with the Blizzard Challenge 2023, the Singing Voice Conversion Challenge (SVCC) \cite{svcc2023}, and the authors of the TMHINT-QI dataset \cite{tmhintqi}, we provided participants with three tracks: Track 1 was French text-to-speech synthesis, Track 2 was singing voice conversion, and Track 3 was noisy and enhanced speech.  All tracks received at least one set of predictions that had a system-level SRCC of over 0.75, demonstrating that zero-shot prediction is challenging (compared to the 0.979 score that could be achieved with a small amount of supervision) but reasonable correlations can still be obtained.

\section{Challenge Description}

The challenge took place from April 28 through June 30, 2023.  The test audio samples for all tracks were released to participants on May 24.  Predictions were due on June 14, and the remaining time was used for compiling and announcing results and collecting teams' system descriptions.

\subsection{Tracks and datasets}

This year's challenge tracks were French text-to-speech synthesis, singing voice conversion, and noisy and enhanced speech.  The details of the datasets will be described next, and statistics of the test phase datasets can be found in Table \ref{tab:data}. 

\setlength{\tabcolsep}{3pt}
\begin{table}
\footnotesize
	\centering
	\caption{Summary of the test phase data for each track.}
	\vspace{5pt}
	\centering
	\begin{tabular}{ c c c c c c}
		\toprule
		Track & Type & Lang & Systems &{\makecell{Samples\\per system}} & {\makecell{\# ratings\\per sample}}\\
		\midrule
		\makecell{Track 1a\\Track 1b} & TTS & Fre & \makecell{Hub: 21\\Spoke: 17} &\makecell{42\\34} & 15 \\
		\midrule
		Track 2 & \makecell{Singing\\VC} & Eng & \makecell{In-dom: 25\\Cross-dom: 24} & 80 & 6 \\
       \midrule
       Track 3 & \makecell{Noisy \& \\ enhanced} & Chi & 97 & 20 & 5.3 \\
		\bottomrule
	\end{tabular}
	\label{tab:data}
\end{table}

\subsubsection{Track 1: French text-to-speech synthesis}

The Blizzard Challenge (BC) is a shared task for text-to-speech synthesis (TTS) that began in 2005 \cite{blizzard2005} and has run for almost every year since.  Participants are provided with a dataset for training TTS systems, which typically consists of high-quality recordings from a professional speaker along with matching transcripts, and they submit generated samples from the synthesis systems that they developed using this data from a script of test sentences.  At the end of the challenge, samples from all teams are rated in an extensive listening test, and the winners are announced.

The Blizzard Challenge 2023 \cite{perrotin23_blizzard} focused on French TTS, with a main Hub task providing 50 hours of training data from a female speaker reading audiobooks, and a speaker adaptation focused Spoke task providing 2 hours from a different female speaker reading a mix of audiobook sentences and transcriptions from French parliament.  18 teams participated, with 14 of those teams participating in the optional Spoke task.  Natural speech from each speaker, as well as samples from two baselines (Tacotron \cite{tacotron} and FastSpeech \cite{fastspeech}) were also considered as systems and evaluated in Blizzard's listening test.  VoiceMOS participants received 42 generated utterances per system for the Hub task and 34 utterances per system for the Spoke task, which were the same samples that were being evaluated in a MOS listening test for quality in parallel with our challenge.  Since two separate listening tests were conducted to evaluate the Hub and Spoke samples, we followed this listening test design for our prediction task and reported the prediction results separately as Track 1a for the Hub task samples and 1b for the Spoke task samples. All audio samples are in 22050 Hz.

\subsubsection{Track 2: Singing voice conversion}

The Voice Conversion Challenge (VCC) is a shared task similar to Blizzard but focusing on speech-to-speech conversion to change the speaker identity \cite{vcc2016, vcc2018, vcc2020}.  This year's challenge targeted singing voice conversion (SVC) and was called the SVCC 2023. There were two tracks for SVCC 2023: Task 1 was in-domain and Task 2 was cross-domain SVC, where the singing voice and speech samples of the target were provided, respectively. Both tracks were in the any-to-one setting, which means only the target speaker/singer datasets were provided, and the systems were asked to convert from unseen singers during evaluation. There were two target singers/speakers (1 male and 1 female) for each of tasks 1 and 2, and there were two source speakers (1 male and 1 female). In total, there were 25 and 24 submitted systems for tasks 1 and 2, respectively, each including two baseline systems. For the listening test, 80 samples were evaluated from each system, and each sample received 6 ratings from a mix of English and Japanese listeners. All audio samples are in 24 kHz.

\subsubsection{Track 3: Noisy and enhanced speech}

Since this is the first time we introduce a noisy and enhanced speech track, we provide an explicit pointer to TMHINT-QI \cite{tmhintqi}, an ``official'' training dataset for this track.  TMHINT-QI is a Mandarin corpus containing 24,408 ten-word samples, including 230 clean utterances spoken by one male and one female speaker, noisy versions generated by with four types of noise (babble, street, pink, and white) at four signal-to-noise ratio (SNR) levels (-2, 0, 2, and 5), and enhanced versions using five speech enhancement (SE) models: KLT, MMSE, FCN, DDAE, and Transformer. 
A total of 226 individuals aged between 20 and 50 took part in the listening test to rate the speech quality of each sample on a range from 1 to 5, with higher values indicating better speech quality.
The subjective scores for each utterance were averaged to derive the ground-truth averaged scores.

A separate dataset, TMHINT-QI(S) \cite{zezario2023}, was created as the test set of Track 3. The noisy utterance generation process was the same as that in TMHINT-QI.
However, the noisy utterances in TMHINT-QI(S) were enhanced using five speech enhancement methods: MMSE, FCN, Trans, DEMUCS, and CMGAN, such that the last two SE methods were different from those used in TMHINT-QI.
A separate, non-overlapping set of 110 listeners (51 male and 59 female) with ages ranging from 20 to 50 and an average age of 32 was recruited, and each listener listened to 100 utterances randomly and evenly selected.
All samples are in 16 kHz.

\subsection{Challenge rules and phases}
\label{ssec:rules}

We ran the challenge on the CodaLab platform\footnote{https://codalab.lisn.upsaclay.fr/competitions/12748}, an open-source web-based platform for machine learning competitions and reproducible research.

Unlike last year's challenge, labeled training data for all tracks was not provided.  However, we provided pointers to several relevant datasets of audio samples with corresponding MOS labels, such as the BVCC dataset \cite{bvcc} which was the training data for last year's challenge, and the SOMOS dataset \cite{somos}, as well as unlabeled datasets with audio in the relevant domains, such as the training data for this year's Blizzard Challenge\cite{bc2023data} and the recommended datasets from SVCC\footnote{http://www.vc-challenge.org/rules.html}.  Participants in BC and SVCC were allowed to also participate in the VoiceMOS Challenge, but they were prohibited from using their own synthesis systems for generating additional data for developing their MOS predictors.

We provided participants with an initial leaderboard phase, during which teams were allowed to make as many submissions as they wanted.  The leaderboard phase used the TMHINT-QI test set, for which training data and labels were already available, and participants could see how well their systems could make predictions and see the results of other teams' submissions.  

For the evaluation phase, we released the audio samples from BC and SVCC, as well as evaluation phase samples for the TMHINT-QI(S) dataset.
Participants were allowed to make up to three submissions for any of the tracks, and no leaderboard was made available as the ground-truth listening test results were also not yet available.  After the end of the evaluation phase, once we received the listening test results from our collaborators, we were able to evaluate each team's submitted predictions and announce the results.

During the post-evaluation phase, we opened a new leaderboard\footnote{https://codalab.lisn.upsaclay.fr/competitions/14273} so that teams could continue to submit predictions and evaluate different system variants.  We also noticed that many teams had not attempted all of the tracks despite the target of the challenge being zero-shot out-of-domain prediction.  So, in an effort to promote more participation in all tracks, we sent participants a message encouraging them to try submitting zero-shot predictions on tracks to which they hadn't previously submitted, using their existing systems.

\section{Participants}

We received evaluation phase submissions from seven teams from academia and three teams from industry for ten in total from seven different countries.  Two teams were returning participants from last year's VoiceMOS Challenge, and three teams indicated that they had participated in similar challenges in the past.  Teams were randomly assigned anonymized numerical team IDs.

We also include results from two of the top systems from last year as baselines: B01 (the SSL-MOS baseline \cite{ssl-mos}), and B02 (the UTMOS system \cite{utmos}).  These systems were not modified in any way or trained on any new data, and their source code and pretrained models were used as-is from their respective open-source repositories\footnote{https://github.com/nii-yamagishilab/mos-finetune-ssl}\footnote{https://github.com/sarulab-speech/UTMOS22}.

\section{Results, Discussion and Analysis}

Only three teams participated in all three tracks initially, and four more teams added predictions for their missing tracks in the post-evaluation phase.  The noisy and enhanced speech track was the most popular, with four teams initially participating in only that track.  

\subsection{Evaluation metrics}

Following last year's challenge, we computed both utterance-level and system-level mean squared error (MSE), linear correlation coefficient (LCC), Spearman rank correlation coefficient (SRCC), and Kendall's Tau rank correlation coefficient (KTAU).  Our primary metric is system-level SRCC with KTAU as a secondary metric, because in a real-life MOS prediction task, we mainly want to know the {\em rankings} of the systems under consideration.  For Tracks 1 and 2, ``systems'' (for the purpose of computing system-level metrics) were considered to be each BC or SVCC team's submission.  For Track 3, systems were considered to be each unique combination of speech enhancement method, SNR, and noise type.

\subsection{Results}

\begin{figure*}[t]
\begin{center}
\includegraphics[width=1.0\textwidth]{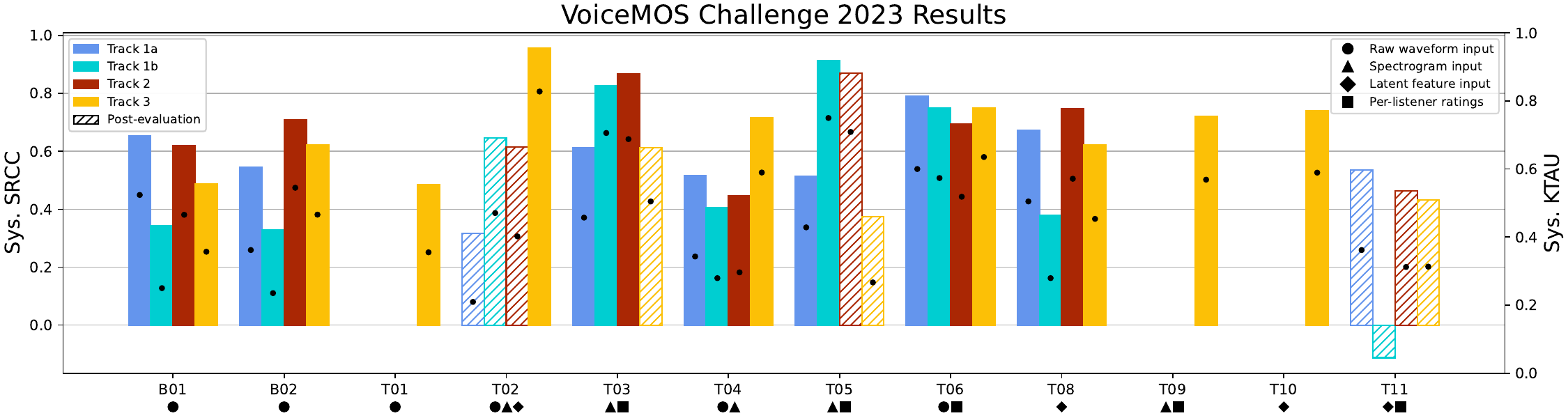}
\end{center}
\vspace{-15pt}
\caption{Results for all tracks of the VoiceMOS Challenge 2023.  Bars indicate system-level SRCC for each track, and dots indicate system-level KTAU.  Hatched bars indicate that the results were received during the post-evaluation phase after the challenge officially ended.  Types of input that each team used are marked under their team ID.}
\label{fig:results}
\end{figure*}

Results are shown in Figure \ref{fig:results}.  The best system-level SRCC score for each track is reported in Table \ref{tab:best}, along with the average system-level SRCC for all teams that participated in each track, and corresponding KTAU values.

\begin{figure*}[hbt!]
\begin{center}
\includegraphics[width=1.0\textwidth]{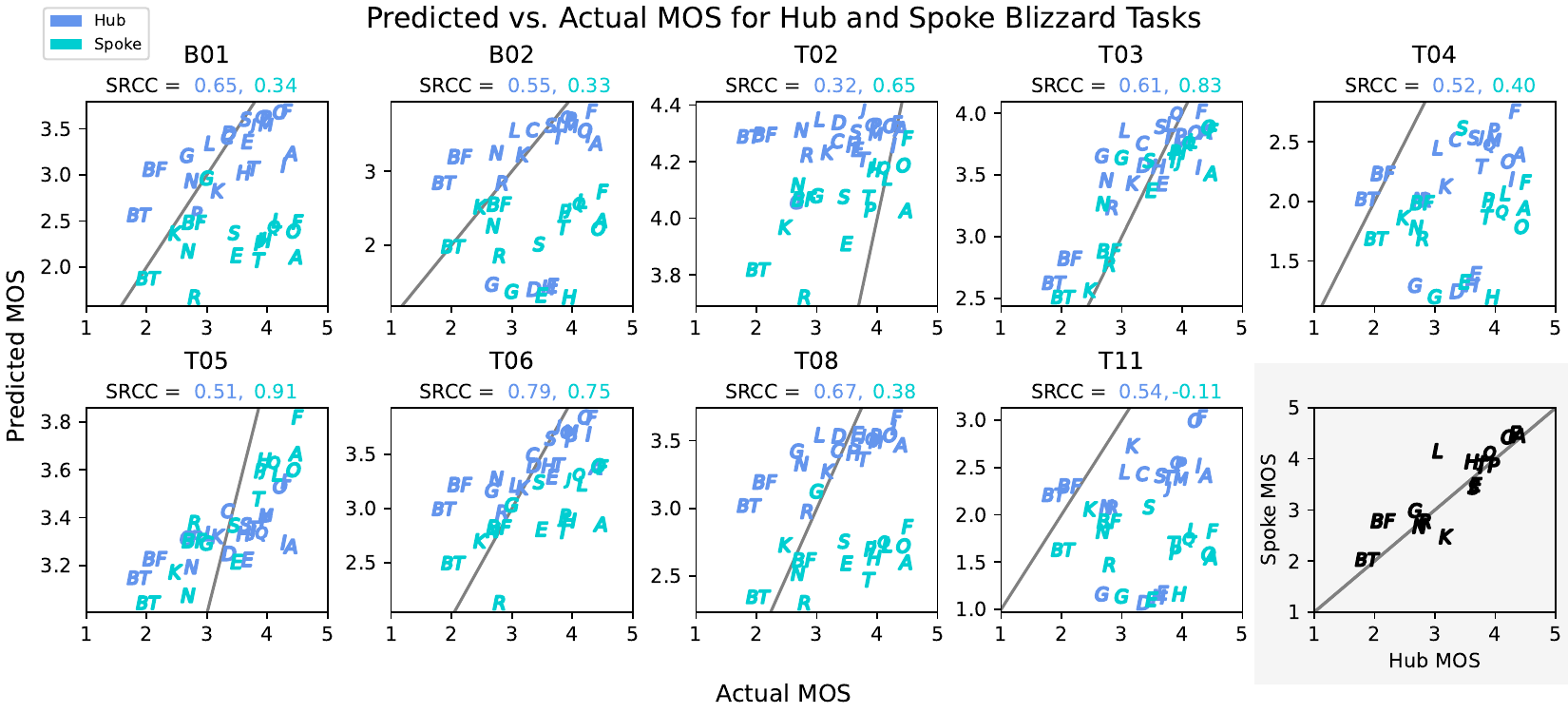}
\end{center}
\vspace{-15pt}
\caption{Predicted vs.\ actual MOS for each VoiceMOS team's prediction of each Blizzard system  for the Blizzard Hub and Spoke tasks (VoiceMOS tracks 1a and 1b, respectively).  Letters indicate Blizzard team IDs.  For reference, the plot in the lower right shows ground-truth hub vs. spoke MOS for each Blizzard team.   Axes have been scaled to the range of each
VoiceMOS team’s predictions for readability, and a $y=x$ line is also shown in each plot to show the skew of the predictions.}
\label{fig:TTS}
\end{figure*}

\setlength{\tabcolsep}{3pt}
\begin{table}

	\centering
	\caption{Best-scoring team (evaluation phase) for each track, along with average SRCC and KTAU for all participating teams for each track.}
	\vspace{5pt}
	\centering
	\begin{tabular}{ c c c c || c c }
		\toprule
		Track & SRCC & KTAU & Team & Avg SRCC & Avg KTAU \\
		\midrule
	1a & 0.79 & 0.60 & T06 & 0.57 & 0.42 \\
    1b & 0.91 & 0.75 & T05 & 0.50 & 0.39 \\
    2  & 0.87 & 0.69 & T03 & 0.67 & 0.50 \\
    3  & 0.95 & 0.83 & T02 & 0.63 & 0.49 \\
		\bottomrule
	\end{tabular}
	\label{tab:best}
\end{table}

It can be observed from Table \ref{tab:best} that the highest overall SRCC score was achieved for Track 3, the noisy and enhanced speech track.  This is not surprising considering that we did provide training and validation data for this track.  The second-highest score was for Track 1b, the speaker-adaptive French TTS track.  Surprisingly, the track with the lowest high score was Track 1a, the basic French TTS task.  The track with the highest average score for teams that participated was Track 2, singing voice conversion.

\subsection{Hub vs. Spoke French TTS}

Interestingly, we can see that for most teams that participated in Track 1, there is a discrepancy in their results between the two sub-tracks.  Both baseline systems had SRCC scores above 0.5 for Track 1a, and below 0.35 for Track 1b.  We can also observe that three teams (T02, T03 and T05) have much higher scores for Track 1b compared to 1a, while two teams (T08 and T11) and two baselines (B01 and B02) have much higher scores for Track 1a compared to 1b.  This was surprising considering that the tasks were not so different in principle.  From listening to the audio samples from each Blizzard task, it seems that the training datasets for each task had somewhat different recording conditions, with the Spoke audio containing reverberation, which may have affected the prediction results.  

We inspected the system-level predictions for Blizzard systems to see whether there were any particular synthesis systems that were difficult for all teams to predict; predicted vs. actual MOS values are shown in Figure \ref{fig:TTS}.

Letters indicate Blizzard team IDs, and BT and BF are the Tacotron and FastSpeech baselines, respectively.  For perfect predictions, we would expect to see Blizzard systems line up along a diagonal line.  We can observe some outliers -- namely, most VoiceMOS teams appear to under-predict Blizzard system R in both tracks; B02 and T04 under-predict Blizzard systems G, D, E, and H; and the natural speech (Blizzard system A) also appears to tend to be under-predicted.  There may be something about these systems that makes them more difficult to predict, but we leave further analysis of this to future work, once more information about these Blizzard systems becomes available.

\subsection{Singing Voice Conversion}

Two teams (T03 and T08) as well as the UTMOS baseline (B02) had singing voice conversion as their best track.  We also observe in Table \ref{tab:best} that the singing voice conversion track was the track with the highest average SRCC for all participating teams.  The ease of predicting MOS for singing synthesized samples came as a surprise, especially since no teams indicated that they used any singing data to develop their systems.

\begin{figure*}[hbt!]
\begin{center}
\includegraphics[width=1.0\textwidth]{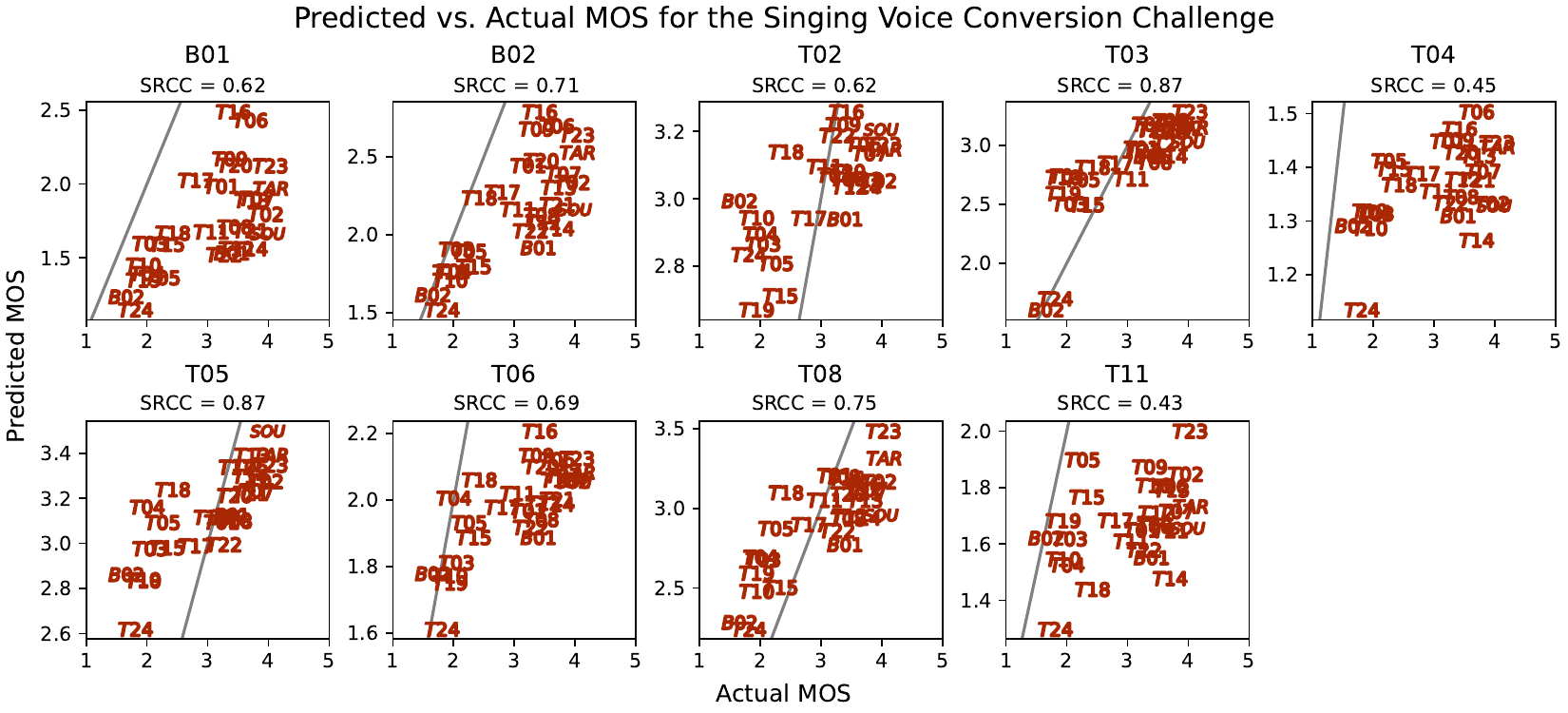}
\end{center}
\vspace{-15pt}
\caption{Predicted vs.\ actual MOS for each VoiceMOS team's prediction of each SVCC system.  Y axes have been scaled to the range of each VoiceMOS team's predictions for readability, and a $y=x$ line is also shown in each plot to show the skew of the predictions.}
\label{fig:svcc}
\end{figure*}

As for systems which were consistently difficult to predict, we observed that SVCC Team T16's system tended to be over-predicted, SOU (source speakers' natural speech) tended to be under-predicted, and Team T24's system also tended to be under-predicted by VoiceMOS teams and baseline systems.  Predicted vs. actual MOS of each SVCC system from each VoiceMOS team is shown in Figure \ref{fig:svcc}.

\subsection{Noisy and Enhanced Speech}

Unsurprisingly, the best overall results were obtained for Track 3, noisy and enhanced speech.  This is most likely due to the availability of in-domain labeled training data, demonstrating that prediction when labeled data from the target domain is available is still an easier task, and zero-shot prediction of a new, unseen domain still has room for improvement.

\begin{figure*}[hbt!]
\begin{center}
\includegraphics[width=1.0\textwidth]{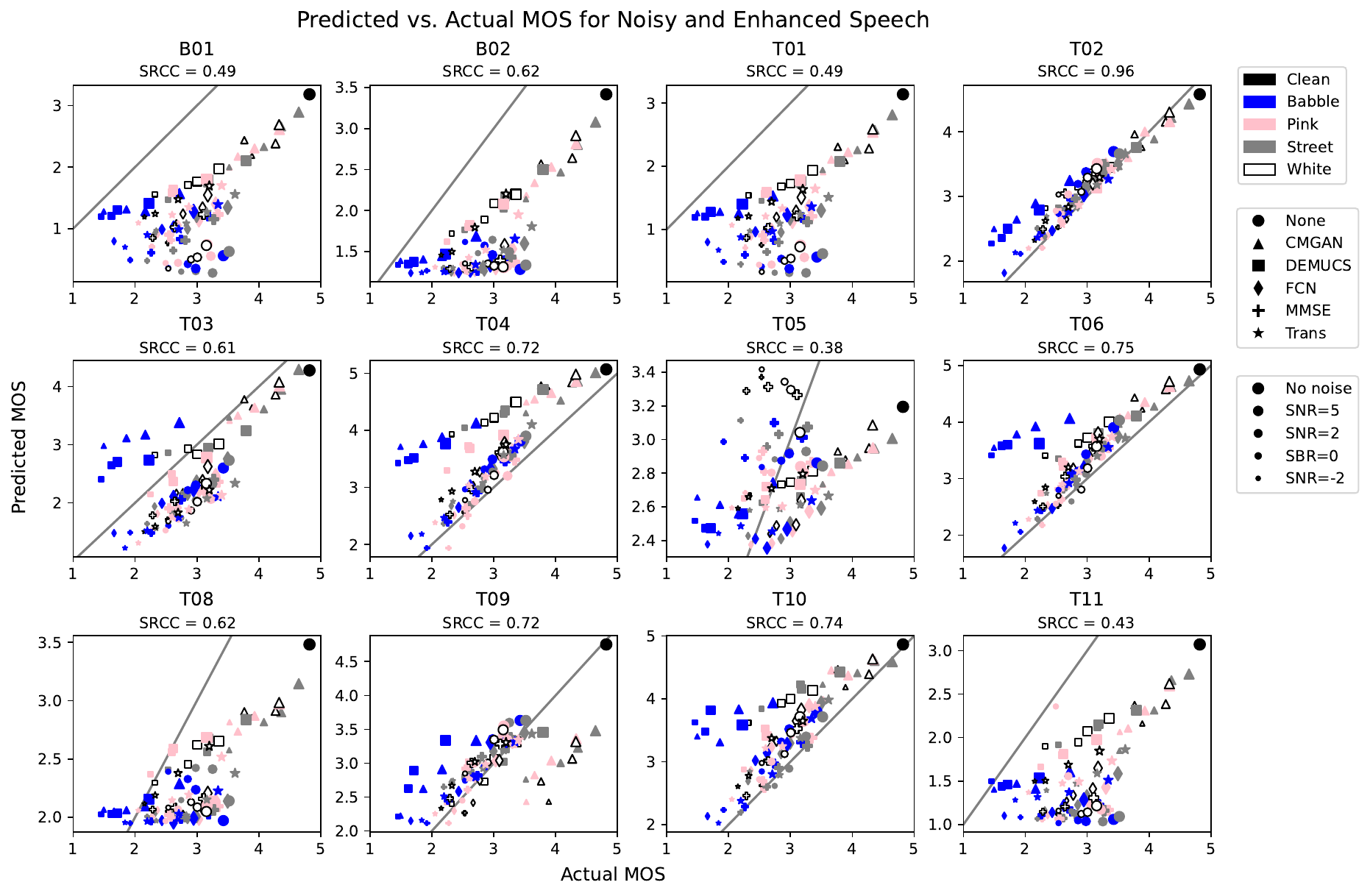}
\end{center}
\vspace{-15pt}
\caption{Predicted vs.\ actual MOS for each VoiceMOS team's prediction of Track 3.  Y axes have been scaled to the range of each VoiceMOS team's predictions for readability, and a $y=x$ line is also shown in each plot to show the skew of the predictions.  Colors show noise type, shape shows speech enhancement system, and the size of the point indicates signal-to-noise ratio.}
\label{fig:track3}
\end{figure*}

Plots of actual vs.\ predicted MOS for Track 3 are shown in Figure \ref{fig:track3}.  It can be observed that the unseen speech enhancement methods, CMGAN and DEMUCS, tend to be over-predicted by several VoiceMOS teams, particularly in the case of babble noise.  Several teams  under-predict CMGAN for the other noise types as well.  This demonstrates that the gap for prediction of MOS for unseen speech enhancement systems still remains even when in-domain training and development data is available.

\section{Analysis of the participating systems}

We received system descriptions from all participating teams.  We asked them to indicate which datasets they used to develop their systems, what kinds of input representations they used, and the details of their model architectures.  We also asked teams to specify what (if anything) they did differently for each track, and whether they developed any part of their approach with special consideration of the zero-shot scenario.

\subsection{Datasets}

Several teams used existing large-scale datasets of MOS-labeled synthesized speech, with four teams using the BVCC dataset, three teams using SOMOS, two teams using past years' Blizzard Challenge data, and one team using the data from VCC 2018.   Five teams indicated that they used the TMHINT-QI data.  No teams indicated that they used any singing datasets in their final submissions.

Teams that used BVCC as the only source of TTS/VC data did not do as well on Tracks 1 and 2 as teams that used a larger variety of sources of TTS/VC data -- this includes the two baseline systems.  T03 did well on Track 2 using extra VCC data from VCC2018, which contains more samples than the VCC2018 samples that are included in BVCC, and the ratings are from a different listening test.   T05 also did well on Track 2 in the post-evaluation phase without using any VC data at all -- they also indicated that they did {\em not} use BVCC, which contains voice conversion samples in addition to TTS, but they did use a variety of TTS MOS datasets in different languages including SOMOS and Blizzard Challenge datasets in Spanish, Mandarin Chinese, and Shanghainese.  This is an interesting result because it indicates that in-domain or even similar-domain data may not always be necessary, and perhaps TTS and VC do not have such large domain differences as we had expected.  The reason may be that many current VC systems use similar components to TTS systems.

\subsection{Differences for Tracks}

Out of the seven teams that participated in all tracks (including during the post-evaluation phase), only one team indicated that they took different approaches for different tracks, stating that they used one model architecture and training dataset for Tracks 1 and 2, and a different model architecture and training data for Track 3.  The other  teams appear to have used the same training data and model architectures for all tracks.  The team that took different approaches for different tracks (T06) appears to have the most consistent performance across tracks, indicating that \textbf{the goal of having one model that can predict MOS well for various different domains has not yet been reached.  }

\subsection{Input Representations}

Six teams used some type of spectrogram-based features as input, and four teams used features generated from other models such as SSL.  
T02 used a mix of different features including latent features from pretrained SSL and ASR models, spectrograms, and learnable filterbank based features, resulting in the best performance on Track 3.

\subsection{Modeling Approaches}

Three teams made modifications to SSL-MOS for their system architecture.  Three teams used LDNet \cite{ldnet} as part of an ensemble, or took a similar approach using per-listener ratings during training.  The teams that used listener information had top results in Tracks 1a, 1b, and 2. This is in contrast to our finding in the last challenge, where we found that using per-listener ratings did not necessary lead to better performance.

We asked teams to report the number of trainable model parameters in their system descriptions, and the smallest number was 768 from T08, who used an SVM-based approach on features extracted from UTMOS.  Interestingly, this team achieved 2nd or 3rd place in Tracks 1a and 2, indicating that simple approaches without training billions of parameters for the target task may still be effective.  It was difficult to observe any pattern in terms of number of trainable model parameters since many teams did not report this information, but our preliminary observations indicate that a larger number of trainable parameters does not necessarily lead to better prediction performance, even when the model is asked to generalize to multiple domains.

We also asked teams whether they took any special considerations for the zero-shot scenario, and only three teams indicated that they did -- one team used concordance correlation in their loss function, one team incorporated unsupervised SpeechLMScore \cite{speechlmscore}, and one team used several Blizzard Challenge datasets in different languages to cover more language variety.

\section{Conclusion}

We have presented the 2023 edition of the VoiceMOS Challenge, which focused on challenging, real-world MOS prediction on different domains of synthesized and processed speech.  The challenge drew ten teams from academia and industry who took a wide variety of approaches to address this difficult zero-shot MOS prediction scenario.  Surprisingly, we found that the domain mismatch between singing voice conversion and text-to-speech synthesis was not as large as we had expected, but we were also surprised at the differences in MOS prediction results between the speaker-dependent and speaker-adaptive French TTS sub-tracks.  We observed that approaches using listener information and a mix of training datasets tended to be more successful.  We can still observe a gap in the ease of making predictions in the case where some labeled in-domain training data is available (Track 3), and the completely zero-shot setting.  Furthermore, no team achieved consistent performance on all tracks using the same model and training data, indicating that \textbf{general-purpose MOS prediction is still an open research problem}.

For future editions of the VoiceMOS Challenge, we aim to close the gap between supervised and zero-shot prediction by encouraging more use of diverse training datasets, including unlabeled ones using unsupervised and semi-supervised approaches. Due to the recent criticism of the MOS paradigm in the speech synthesis community, we are also considering to shift to prediction of other subjective tests such as preference tests.

\section{Acknowledgments}

This study was supported by JST CREST Grant Number JPMJCR18A6 and JPMJCR20D3, MEXT KAKENHI grant 21K11951, and MOST 110-2221-E-001-015-MY3.  This work was also partly supported by JSPS KAKENHI Grant Number 21J20920, and JST CREST Grant Number JPMJCR19A3. We would like to thank Olivier Perrotin and the organizers of the Blizzard Challenge for making their samples and listening test results available for our challenge and for answering questions and providing information about the listening test, the organizers of the Singing Voice Conversion Challenge for making their samples and listening test results available for our challenge, and Yu-Wen Chen for answering questions about TMHINT-QI.

\clearpage

\bibliographystyle{IEEEbib}
\bibliography{refs}

\end{document}